\documentclass[aps,prl,twocolumn,superscriptaddress]{revtex4-1}
\usepackage{graphicx}  % needed for figures
\usepackage{amsmath} % needed for splitting equations into lines

\usepackage{graphicx}
\usepackage{amsmath,amssymb,amsfonts}
\usepackage{textcomp}
\usepackage{hyperref}
\usepackage{gensymb}
\usepackage{verbatim}

\hypersetup{breaklinks=true,colorlinks=true,urlcolor=black}
\usepackage{color}
\usepackage{soul}
\usepackage{gensymb}

\DeclareGraphicsExtensions{.jpg,.pdf,.png,.eps}

\begin{document}

% Use the \preprint command to place your local institutional report
% number in the upper righthand corner of the title page in preprint mode.
% Multiple \preprint commands are allowed.
% Use the 'preprintnumbers' class option to override journal defaults
% to display numbers if necessary
%\preprint{}

%Title of paper
\title{Construction of $A$-$B$ hetero-layer intermetallic crystals: case studies of the 1144-phase TM-phosphides \textit{AB}(TM)$_4$P$_4$ (TM=Fe, Ru, Co, Ni)}

% repeat the \author .. \affiliation  etc. as needed
% \email, \thanks, \homepage, \altaffiliation all apply to the current
% author. Explanatory text should go in the []'s, actual e-mail
% address or url should go in the {}'s for \email and \homepage.
% Please use the appropriate macro foreach each type of information

% \affiliation command applies to all authors since the last
% \affiliation command. The \affiliation command should follow the
% other information
% \affiliation can be followed by \email, \homepage, \thanks as well.

\author{B. Q. Song}
\author{Mingyu Xu}
\affiliation{Ames Laboratory, Iowa State University, Ames, Iowa 50011, USA}
\affiliation{Department of Physics and Astronomy, Iowa State University, Ames, Iowa 50011, USA}
\author{Vladislav Borisov}
\affiliation{Institute of Theoretical Physics, Goethe University Frankfurt am Main, D-60438 Frankfurt am Main, Germany}
\affiliation{Department of Physics and Astronomy, Uppsala University, 752 36 Uppsala, Sweden}
\author{Olena Palasyuk}
\affiliation{Ames Laboratory, Iowa State University, Ames, Iowa 50011, USA}
\author{C. Z. Wang}
\affiliation{Ames Laboratory, Iowa State University, Ames, Iowa 50011, USA}
\affiliation{Department of Physics and Astronomy, Iowa State University, Ames, Iowa 50011, USA}
\author{Roser Valent\'\i}
\affiliation{Institute of Theoretical Physics, Goethe University Frankfurt am Main, D-60438 Frankfurt am Main, Germany}
\author{Paul C. Canfield}
\affiliation{Ames Laboratory, Iowa State University, Ames, Iowa 50011, USA}
\affiliation{Department of Physics and Astronomy, Iowa State University, Ames, Iowa 50011, USA}
\author{K. M. Ho}
\affiliation{Ames Laboratory, Iowa State University, Ames, Iowa 50011, USA}
\affiliation{Department of Physics and Astronomy, Iowa State University, Ames, Iowa 50011, USA}

%Collaboration name if desired (requires use of superscriptaddress
%option in \documentclass). \noaffiliation is required (may also be
%used with the \author command).
%\collaboration can be followed by \email, \homepage, \thanks as well.
%\collaboration{}
%\noaffiliation

\date{\today}

\begin{abstract}
The discovery of the 1144-phase, e.g. CaKFe$_4$As$_4$, creates opportunities to build novel intermetallics with alternative stacking of two parent compounds. Here we formalize the idea by defining a class of bulk crystalline solids with $A$-$B$ stacking (including 1144-phases and beyond), which is a generalization of hetero-structures from few-layer or thin-film semi-conductors to bulk intermetallics. Theoretically, four families of phosphides \textit{AB}(TM)$_4$P$_4$ (TM=Fe, Ru, Co, Ni) are investigated by first-principles calculations, wherein configurational, vibrational and electronic degrees of freedom are considered. It predicts a variety of stable 1144-phases (especially Ru- and Fe-phosphides). Stability rules are found and structural/electronic properties are discussed. Experimentally, we synthesize high-purity CaKRu$_4$P$_4$ as a proof of principle example. The synthetic method is simple and easily applied. Moreover, it alludes to a strategy to explore complex multi-component compounds, facilitated by a phase diagram coordinated by collective descriptors. 
\end{abstract}

% insert suggested PACS numbers in braces on next line
% insert suggested keywords - APS authors don't need to do this
%\keywords{}

%\maketitle must follow title, authors, abstract, \pacs, and \keywords

% body of paper here - Use proper section commands
% References should be done using the \cite, \ref, and \label commands
\maketitle

 \section{1. Introduction}
Breakthroughs in synthesis could evoke new material concepts. For example, graphene together with a variety of single-layer structures have substantiated and popularized the 2-D materials \cite{Geim}. On the other hand, varying the attributes affiliated with existing conceptions could help discover new types of compounds. For example, the finding of high-entropy alloys (HEA) \cite{HEA04, HEA} was motivated by the idea of tuning the number of alloy constituents and concentrations, which are two basic attributes based on present concepts of alloys. Such synthesis-conception duality is essential for material development, and the dual aspects are mutually inspiring. The recent discovery and study of $AeA$Fe$_4$As$_4$ ($A$ = alkali, $Ae$ = alkaline earth) 1144-phase compounds \cite{Iyo, Mou, MeierPRM,MeierNat, Song, pinSC, bilayer, 2dSC} is such an example. Its synthesis implied creating quaternary compounds by judicious combinations of related ternary materials $A$Fe$_2$As$_2$ and $Ae$Fe$_2$As$_2$; on the other hand, choosing and varying the cation attributes can yield broader intermetallic compounds, as conceptualized in this work.
\begin{figure}
\includegraphics[scale=0.7]{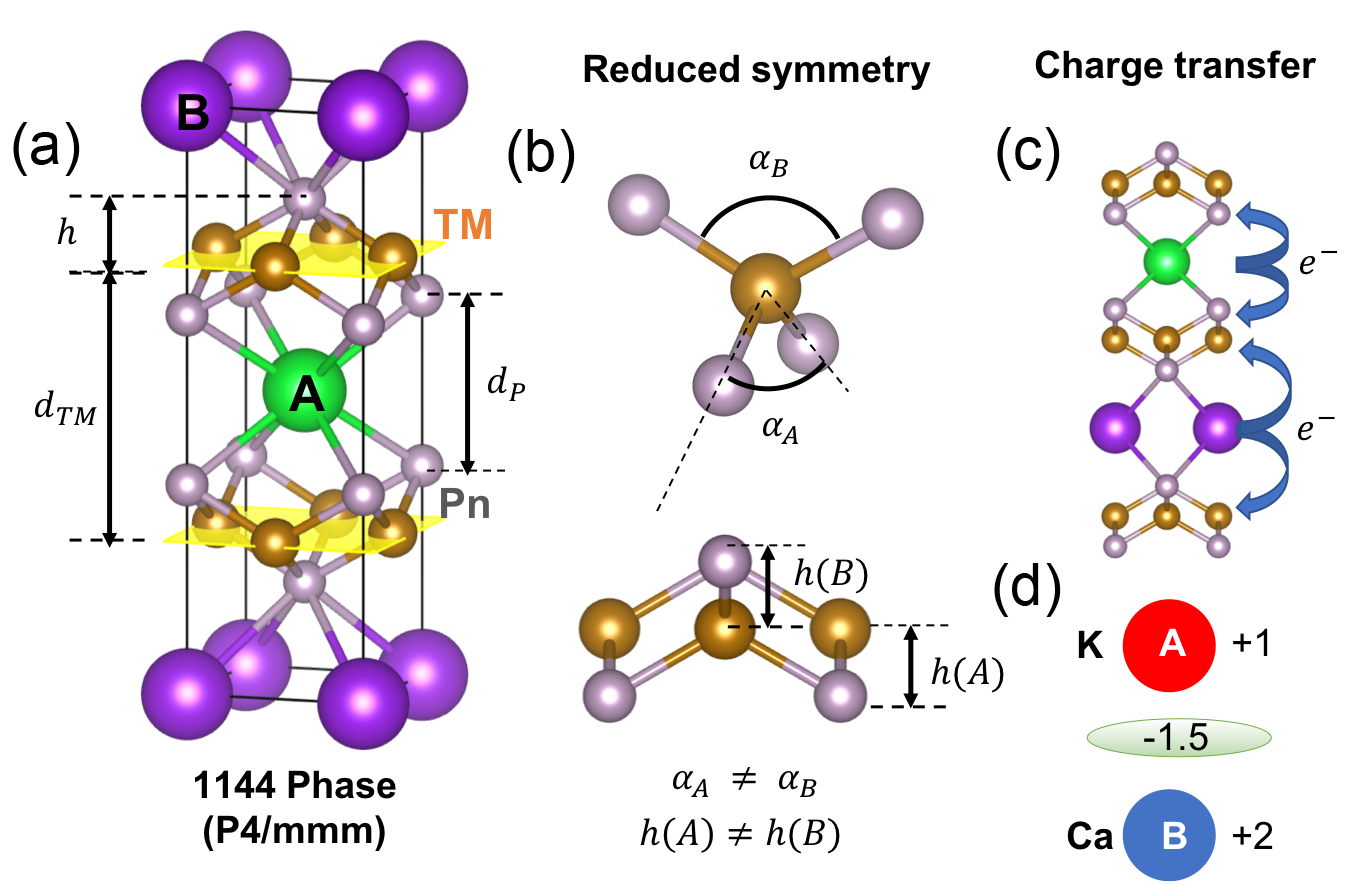}
\caption{\label{fig:epsart}(color online): (a) The crystal structure of the 1144-phase. If \textit{A}, \textit{B} sites are occupied by the same cations, it becomes the 122-phase (b) Structure parameters with broken glide symmetry. (c) Charge transfer from cation-layers to TM-layers. (d) Schematic charge distribution of KCaRu$_4$P$_4$. The layers of K$^+$ and Ca$^{2+}$ take positive charges; each TM-layer takes -1.5e per unit cell. Pn stands for pnictogen and TM stands for transition metals\label{f1}}
\end{figure}

The 1144-phase (Fig.~\ref{f1}(a)) can be regarded as alternating cation layers in the 122-phase being replaced by hetero-cations \cite{Iyo}. Stoichiometrically, it is equivalent to 50\% doping of \textit{Ae}Fe$_2$As$_2$ by \textit{A} (or \textit{A} by \textit{Ae}). However, this view obscures that Fe-As layers have been significantly altered due to cation polarization in 1144-phases, which sets it apart from 122-phases \cite{Sasmal, Ralloy, SrNa}. Compared with 122-phases, 1144-phases feature distinct parameters (e.g., bonding angles, pnictogen heights) above and below the transition metal (TM) layer (Fig.~\ref{f1}(b)). Consequently, the \textit{n}-glide plane across the Fe-layer disappears, and the space group is reduced from $I4$/$mmm$ in 122-phases, to $P4$/$mmm$ in 1144-phases \cite{Song}. This fact might be relevant to the emergence of hedgehog spin textures \cite{MeierNat}, low-dimensional superconductivity (SC) \cite{2dSC}, FM-SC coexistence \cite{CaoRbEu}, suppressed $T_c$ with pressure \cite{Xiang}, and disputed nematic behavior \cite{MeierNat, NoNeum, Anna_Boehmer, SrNa}. Besides, the 1144 crystal surface has been suggested to be a platform for testing the Majorana zero modes \cite{DingH, Surf}.

The 1144-phase is a representative of “ordered stacking” \cite{ChgCao}, and poses an intriguing conjecture that such compounds widely exist subject to certain rules about building blocks: cation- and TM-layers. Alloys mix elementary atoms into random solid solution (SS), while 1144-phases have layers as basic units and pack them into an ordered pattern. Since interesting phenomena (e.g., SC \cite{MeierNat, pinSC, 2dSC}, magnetism \cite{CaoRbEu}) arise from \textit{d}-orbitals in TM-layers, the most revealing view is that the TM-layer resides in between two hetero-cation layers (Fig.~\ref{f1}(a)). It is reminiscent of 2-D materials being sandwiched by the substrate and vacuum. Varying species of hetero-cations serves as attributes to adjust chemical potentials and symmetry, just like choosing an underneath substrate and on-top dopants as the counterpart tuning for graphene \cite{Geim, Gap}. The pnictogen height \textit{h} (Fig.~\ref{f1}(a)) exhibits considerable asymmetry on different sides with 5\%-10\% changes compared with 122-phases. The manipulation could be further enriched by the diverse candidates for substitution: Ru-Si \cite{RRuSi}, VO$_2$\cite{VO2} for TM-layers; and \textit{A}, \textit{Ae}, rare earth (RE) for cation layers \cite{bqs}. 

The paper is organized as follows. Sec. 2 introduces a general class of intermetallics inspired by the 1144-phase. In Sec. 3, four families of phosphides \textit{AB}(TM)$_4$P$_4$ (TM=Fe, Ru, Co, Ni) are chosen for a case study. We discuss their formation conditions at zero and finite temperatures, accounting for the configurational, vibrational, and electronic degrees of freedom. A number of 1144-compounds (especially Ru- and Fe-based phosphides) are predicted to be stable. Other questions are also examined regarding the stability mechanism, electronic structures, and latent features of the 1144 crystals. Sec. 4 presents our work of synthesizing high-purity KCaRu$_4$P$_4$. Its structure is determined as a proof of principle example. %measured to validate the prediction. 
In Sec. 5, we survey that 1144-like crystals are broadly existent, tuning-friendly, and of versatile physical properties, such as interplay of magnetism and superconductivity \cite{David, DaiP}, volume-collapsed phases \cite{Collapse, Borisov, TwoTr}, or heavy fermion behaviors \cite{RRuSi}. It also presents an avenue for designing complex many-constituent intermetallic compounds beyond ternary alloys.
%\begin{figure}
%\includegraphics[scale=0.37]{figure2_test.png}
%\caption{\label{fig:epsart}(color online): HC is an effective binary system for cations $A$, $B$ ($N$=2). At small $N$, SS (blue area) tends to form; at large $N$, phase separation (green) is easy to take place. HC and HEA are two abnormal states of matter around an equiatomic concentration (dashed line), which resist the tendency in their own regimes. Thus, HC (HEA) require special formation conditions and appear as "islands" in the blue (green) region. Inset: schematics for HC and SS, where colors stand for cation occupancy and blended colors mean 50\% probability. HC display distinct features from SS, regarding symmetries, spin states, etc. Transition between HC and SS can be characterized by asymmetry parameter $\varepsilon$ (defined in the main text); HC is not a single point, but exists in a range of $\varepsilon$, which provides tolerance for imperfection. \label{f2}}
%\end{figure}
\section{2. Definition of hetero-layer intermetallic crystals}

The 1144-phase raises many questions. Why an ordered phase emerges at 50\% concentration, where configuration entropy gets to maximum? This is utterly opposite to high entropy alloys \cite{HEA}, which target the equal-concentration for a maximum tendency to randomness. Whether such ordered compounds exist beyond Fe-arsenides, or beyond 1144-phases? If so, what are their common features? Which minor distinctions can be ignored? Is there a general descriptor for them?  

Working on these questions, we naturally obtain a broader class of $A$-$B$ hetero-layer intermetallic crystals defined below. For short, we call it hetero-crystal: ``hetero" refers to $A$-$B$ stacking, ``crystal" reminds that it is bulk rather than low-dimension or thin film. Note that hetero-crystal is merely an abbreviation in this context. 

(i) They are bulk crystals (i.e., stacking units are atomic thick and periodic along the $z$-axis), manufactured by liquid growth \cite{MeierPRM} or solid state reaction \cite{Iyo}, in contrast to low-dimensional or thin-film hetero-structures (like super-lattice \cite{Superlat} or tunneling junctions \cite{Hstr}), which are usually prepared with vapor deposition. 

(ii) They consist of two sub-systems: cation layers (usually metal elements) and skeleton layers (e.g., TM-pnictogen or TM-chalcogen layers). Note that ``layer" is indispensible, i.e., it requires bonding along the $z$-axis is weaker than that in the $x$-$y$ plane. Thus, structures without layer divisions, such as pyrochlores and fluorite, are not within the present definition, although it might exhibit cation ordering \cite{PFtr, NaYbO2}. The layer construction significantly affect physical properties. For example, the two sub-systems differ in electronegativity and charge transfer between them (Fig.~\ref{f1}(c)(d)) leads to an ionic type of inter-layer bonding. The TM-layer motif (usually tetrahedron) causes particular hybridizations of \textit{d}-orbitals that generate partially filled bands, promoting metallic properties and itinerant magnetism \cite{David}. These features set it apart from hetero-structures in the context of semiconductors \cite{Hstr}. 

(iii) They show an \textit{A}-\textit{B} stacking of alternating cation layers. As such, the concentration of \textit{A} (or \textit{B}) is fixed to 50\% and disorder in cation layers is much suppressed. More importantly, the \textit{A}-\textit{B} stacking creates asymmetric up-and-down environments and distorts the TM-layer, which could be Fe-Pn, NiO$_2$ \cite{NiX, NiL}, VO$_2$, and Mo$_N$O$_{3N-1}$ ($N$ =1, 2, 3...) \cite{VO2}. The structural change associated with the symmetry breaking is huge compared to that by applying pressure. For instance, our calculation shows that the asymmetry for pnictogen height $h$ (Fig.~\ref{f1}(b)) is up to 5\%${\sim}$10\%. Thus it seems plausible that HC will behave distinctively from SS phases with identical stoichiometry. 

(iv) They are formed by combining two \textit{stable} parent phases. For example, the 1144-phase can be synthesized by mixing two 122-compounds, as done in this work. In general, the parent phase refers to the phase that has the common skeleton layer as hetero-crystals, but has mono-cations. It requires that the involved parent phases are stable (or meta-stable) \cite{bqs}. (iv) suggests that seeking HC should begin with looking for stable parent phases. Hundreds of 122-phases have been synthesized \cite{122Base}, and the reservoir of parent compounds facilitate synthesis.

Besides the four definitions above, it is worth mentioning that although the 1144-phase is quaternary, it is effectively binary, as the main variables are the cations $A$ and $B$. When the parent phases are mixed, entropy favors random mixing of $A$ and $B$, forming a uniform SS phase; whereas if an ordered phase occurs, it must have been favored by enthalpy. Thus, the 1144-phase's emergence is closely linked to enthalpy battling with configuration entropy. %At equiatomic concentration, the enthalpy approximately scales as $N^2$ ($N$ is the number of constituents), while entropy scales with log$N$ \cite{Duane}. That is, for smaller $N$, SS phases tend to dominate, as commonly seen in binary alloys. For large $N$, ordered states will form. In plain words, the more ingredients being mixed, the more difficult it is to form a uniform ``soup", and the mixture tends to undergo phase separation (PS). Thus, the occurrence of ordered HC ($N=2$) and disordered HEA ($N{\geq}5$) are both abnormalities (Fig.~\ref{f2}); thus HC should require special formation conditions. In the following sections, we shall discover these conditions, and evaluate entropy and enthalpy in the context of 1144-phases. 
\begin{figure}
\includegraphics[scale=0.46]{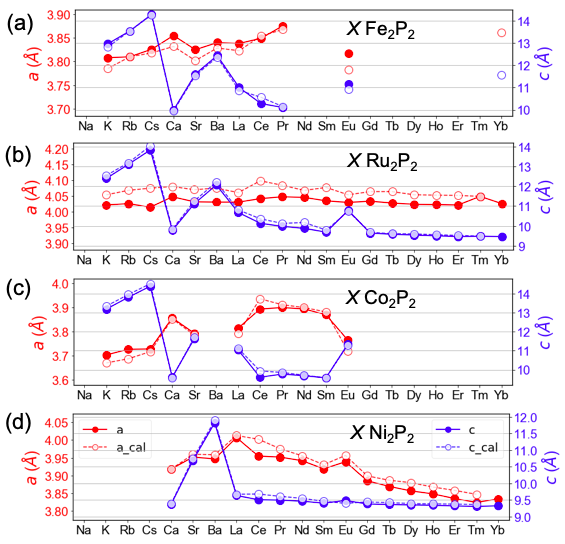}
\caption{\label{fig:epsart}(Color online): Lattice-parameters $a$ (red) and $c$ (blue) of the tetragonal unit of 122-phases. Filled and open points represent experimental and theoretical values. Missing points are unstable or not reported yet. Experimental results are cited from Springer Materials data base.\label{f3}}
\end{figure}

\section{3. Theoretical results}
\subsection{3A. Stability of 1144-phase}
The 1144-phase is presently limited to Fe-arsenides with cations from the alkali metal group (IA) or alkaline earth group (IIA) \cite{bqs}, and it is highly desirable to extend the chemical scope. Empirical rules for stable 1144-phases have been recognized: a large cation-radius mismatch ${\Delta}R$ seems a favorable factor; besides, the lattice mismatch ${\Delta}a$ is also relevant \cite{Iyo, MeierPRM}. The dependence of the two descriptors ${\Delta}R$ and ${\Delta}a$ can be rationalized with the enthalpy change due to elastic deformations \cite{Song}. The mechanism is generic, so it applies to Ru-P 1144 systems and other TM-pnictides. On the other hand, descriptors ${\Delta}R$ and ${\Delta}a$ fail to encode charge information (i.e., valence states exhibited by the cation) \cite{ChgCao}, thus are incapable of treating situations with valence variation, e.g. when trivalent cations rare earth RE$^{3+}$ are introduced. This implies additional descriptors are needed.

We start the search from parent compounds of 122-phase pnictides, which are broadly existing for different valences ($X=$Ai$^+$, Ae$^{2+}$, or RE$^{3+}$), an ideal ground for sheding light on the charge effect. Also, the iso-valence substituion P$\rightarrow$As is likely to yield stable phases. In terms of properties, Ru might introduce strong spin-orbital coupling; radius of P is smaller, thus it is easier to realize the half-collapsed phase \cite{Collapse, Borisov, TwoTr}. The lattice parameters (experimental and calculated) are listed in Fig.~\ref{f3}(a)-(d). Notably, 122-phases are unstable with certain cations. The calculation is based on density functional theory (DFT) \cite{Perdew} with Perdew Burke Ernzerhof (PBE) exchange-correlation functional, implemented by the Vienna ab initio simulation package (VASP).  Both the unit cell and internal coordinates are fully relaxed. Calculation parameters are shown in the supplementary information (SI) \cite{SI}. The results of the calculated lattice parameters are close to the experimental values for most of systems with differences  $\leq$1\%. Except for Ce-contained 122 phases where lattice parameters are consistently overestimated, probably due to inherent issues of pseudo-potentials in treating variant valences of Ce.  

The 1144-phase is achieved by mixing two 122-phases with common skeleton layers (choose from the four groups in Fig.~\ref{f3}). The reaction merely involves redistribution of cations, either ordered or random. Thus, the question becomes the competition between 1144-phase and 122-solution-phase (122(s)-phase) \cite{Iyo, Song}. In general, the simplifying to two-phase competition is made possible by condition (iv) defined in Sec. 2, i.e., the existence of stable parent phases. This is indispensable, because having stable parent phases provides starting points and allows the reaction to be studied in a simpler landscape.

\begin{figure}
\includegraphics[scale=0.52]{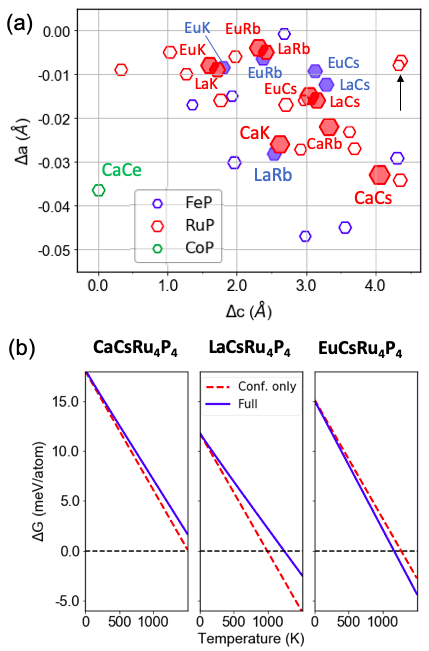}
\caption{\label{fig:epsart}(Color online): (a) ${\Delta}H$ (proportional to the hexagon size) plotted against two descriptors: ${\Delta}a = -{\mid}a_{122}^{A}-a_{122}^{B}{\mid}$, ${\Delta}c = {\mid}c_{122}^{A} - c_{122}^{B}{\mid}$, where $a$ and $c$ are lattice parameters of the two parent 122-phases \textit{A}(TM)$_2$P$_2$ and \textit{B}(TM)$_2$P$_2$. The most promising systems are highlighted with solid hexagons and the corresponding cations are denoted. (b) Temperature dependence of free energy ${\Delta}G=G_{122(s)}-G_{1144}$. (${\Delta}G>0$ 1144-phase is stable.) The dashed line only includes $S_{conf}$. The blue line further includes $S_{vib}$ and $S_{e}$.\label{f4}}
\end{figure}

At zero temperature, phase stability is dictated by the formation enthalpy ${\Delta}H = H_{122(s)} - H_{1144}$. The enthalpy of the ordered phase can be estimated from super-cell calculations within DFT. For the solution phase, however, calculation of enthalpy involves random configurations and we adopt the ideal solution approximation.
\begin{equation}
H_{122(s)}=x{\cdot}H_{122}^{A}+(1-x){\cdot}H_{122}^{B},
\label{Hx}
\end{equation}
where $x$ is the concentration of cation $A$, and $x=$1/2 in this case. The stable 1144-phases (${\Delta}H{>}5$ meV/atom) are summarized in Fig.~\ref{f4}(a) (complete list seen in Sec. 2 of SI \cite{SI}) where
${\Delta}H$ is plotted against two descriptors:
\begin{equation}
{\Delta}a = -{\mid}a_{122}^{A}-a_{122}^{B}{\mid},
\label{descriptor_a}
\end{equation}
\begin{equation}
{\Delta}c = {\mid}c_{122}^{A} - c_{122}^{B}{\mid}
\label{descriptor_c}
\end{equation}
with $a$ and $c$ denoting the lattice parameters of the tetragonal unit cell of parent 122-phases. A number of 1144-phases are found stable, especially for Fe- and Ru-phosphides. The stable 1144-phases concentrate in a region of small $|{\Delta}a|$ and large $|{\Delta}c|$ (upper right corner of Fig.~\ref{f4}(a)), a tendency aligned with earlier findings \cite{Iyo, Song}. However, a closer inspection shows some discrepancies. For example, the two red hexagons (black arrow) deep in the upper-right corner (CsErRu$_4$P$_4$ and CsHoRu$_4$P$_4$) are not the most stable albeit they best satisfy the criteria. This implies additional descriptors, as discovered shortly.

At finite temperature, the phase stability is determined by the free energy:
\begin{equation}
\begin{split}
{\Delta}G&=G_{122(s)}-G_{1144} \\
&={\Delta}H+{\Delta}E_0-{\Delta}S_{conf}T-{\Delta}S_{vib}T-{\Delta}S_{e}T,
\end{split}
\label{eq4}
\end{equation}
where $E_0$ is the zero-point vibration energy. $S_{conf}$, $S_{vib}$ and $S_e$ are configurational, vibrational and electronic entropies, respectively. The configuration entropy $S_{conf}$ for the 1144-phase is zero. The $S_{conf}$ of the 122(s)-phase (per unit cell) is estimated by \cite{Gaskell}
\begin{equation}
S_{conf}=k_{B}(x{\log}x+(1-x){\log}(1-x)),
\label{eq5}
\end{equation}
where \textit{x} is defined in Eq.~\ref{Hx}. In this case, $S_{conf}$  is a constant 0.012 meV/(atom K). The electron entropy is estimated by
\begin{equation}
S_{e}=-k_{B}{\int}D(E)({f{\cdot}{\log}f+(1-f){\log}(1-f)}{\cdot})dE,
\label{eq6}
\end{equation}
where $f$ is the Fermi distribution function and $D(E)$ is the density of states, which are obtained from DFT calculations. Calculations of zero-point energy and vibration entropy $S_{vib}$ are performed by the code \textit{phonopy} within the harmonic approximation \cite{Phonon}. For the 122(s)-phase, $S_{vib}$ and $S_{e}$ are estimated with the average of two 122-phases, similar to Eq.~\ref{Hx}.

In Fig.~\ref{f4}(b) we plot ${\Delta}G$ as a function of temperature for three typical stable systems. At high temperature, the entropy will eventually bring the ordered 1144-phase into the 122(s)-phase. We define $T^{*}$ as the temperature where ${\Delta}G$=0 and, below which, the 1144-phase is stable. Comparing the dashed (only includes $S_{conf}$) and solid curves in Fig.~\ref{f4}(b), we find that including $S_{vib}$ and $S_{e}$ modifies minimally the free energy, and thus conclude that the $S_{conf}$ is the main contribution, which is consistent with previous findings \cite{Song}. As such, ${\Delta}H/ S_{conf}$ may give a rough estimate of $T^{*}$. In Sec. 3 of \cite{SI} we provide ${\Delta}G$ calculations for more systems.
 \begin{figure*}
\includegraphics[scale=0.43]{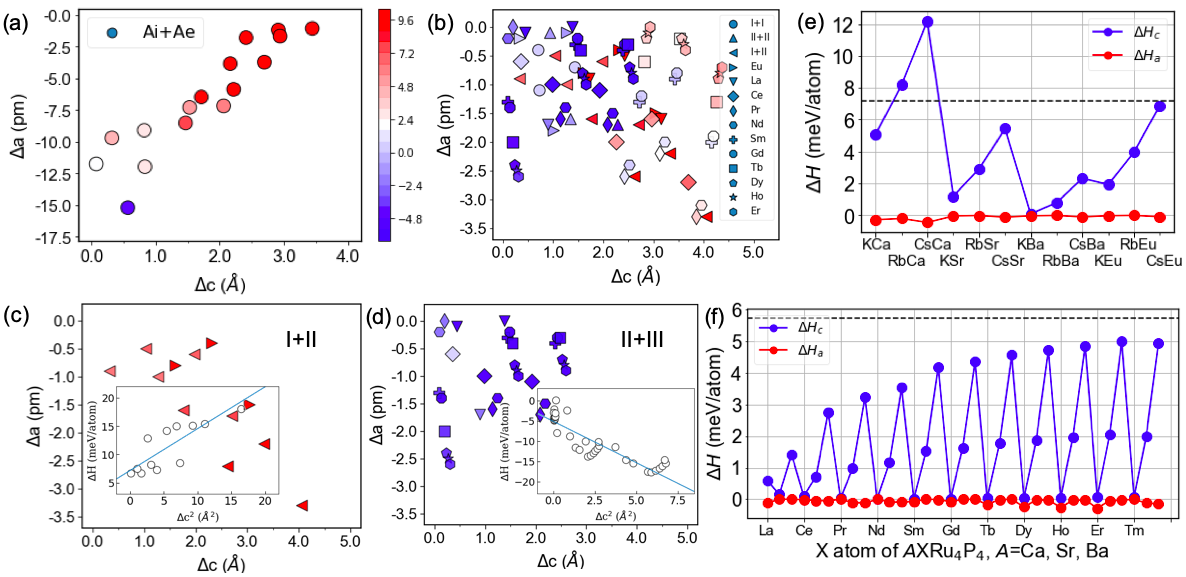}
\caption{\label{fig:epsart}(Color online): The formation enthalpy ${\Delta}H$ of (a) \textit{AB}Fe$_4$As$_4$ with $A$ $B$ being alkali metal (Ai) and Alkaline earth (Ae), (b) of \textit{AB}Ru$_4$P$_4$. Each spot represents a particular pair of $A$-$B$. Color scales represent ${\Delta}H$, and red means stable 1144-phase (unit: meV/atom). I+II means alkali metal (AI) plus alkaline earth (AII); II+II means AII+AII; I+II means AI+AII. Eu, La, etc., mean the specific rare earth (RE) combined with another AI or AII. The ${\Delta}H$ of subgroups (c) I+II and (d) II+III. Other subgroups are shown in \cite{SI}. The shapes of points are consistent with the legend in (b). (e)(f) ${\Delta}H$ contribution from ${\Delta}H_a$, ${\Delta}H_c$ and ${\Delta}H_0$. The dashed line represents the magnitude $|{\Delta}H_0 |$. Note that ${\Delta}H_0$ is positive for (e), but negative for (f). \label{f5}}
\end{figure*}

\subsection{3B. General rules of phase stability}
The stability of 1144-phases obeys certain rules. As shown previously, descriptors ${\Delta}c$ given by Eq.~\ref{descriptor_c} (or atomic radius mismatch ${\Delta}R$), and ${\Delta}a$ given by Eq.~\ref{descriptor_a}, proved useful \cite{Iyo,Song}. They were first exemplified in Fe-arsenide 1144-systems and understood
with an elastic model \cite{Song}. Next we will find a descriptor associated with the charge balance between cation layers and pnictogen (Pn) layers, which will subtly affect the 1144/122(s) stability.

The formation enthalpies ${\Delta}H$ of $AB$Fe$_4$As$_4$ are plotted in Fig.~\ref{f5}(a), which shows that the stability gets enhanced from the lower-left to the upper-right in the graph. Thus, $AB$Fe$_4$As$_4$ endorse the established rule \cite{Iyo, Song} that the 1144-phase is stabilized with large $|{\Delta}c|$ and small $|{\Delta}a|$. However, the rule seems inadequate for Ru-phosphides (Fig.~\ref{f5}(b)), as the stable and unstable systems remain unseparated. To elucidate the situation, we divide these data into subgroups by cation valence states, e.g., I + II (Fig.~\ref{f5}(c)) and II + III (Fig.~\ref{f5}(d)); then the rule manifests itself. Basically, the cation pair affects the stability: group I+II form a stable 1144-phase and II+III do not. Such distinction, obviously, relies on charge balance between the cation layer and TM-layer. In fact, charge balance is a generic factor, which also affects the 122-phases (Fig.~\ref{f3}(a)-(d)). For instance, mono-valence cations do not favor the 122-phase $A$Ni$_2$P$_2$ (Fig.~\ref{f3}(d)). On the other hand, tri-valence cations never stabilize $A$Fe$_2$As$_2$ \cite{bqs}. Note that the favorable charge balance for Ru-phosphides is alkali metal (AI) + alkaline earth (AII) (or AI+Eu), i.e., the effective valence state 1.5+, which is the same as in the Fe-arsenide \cite{Song, bqs} systems, consistent with the anticipation of iso-valence substitution.

At the same time, the size effect still plays a role, as ${\Delta}H$ manifests a linear correlation with $({\Delta}c)^2$ in each subgroup (insets of Fig.~\ref{f5}(c) and (d)). It suggests that the charge effect cooperates with the size effect, for which we formulate the formation enthalpy ${\Delta}H$ as two additive parts regarding the size and charge effects,
\begin{equation}
{\Delta}H={\Delta}H_{size}+{\Delta}H_{charge}
\label{eq7}
\end{equation}

The size part essentially accounts for the elastic energy and could be approximated as \cite{Song}
\begin{equation}
{\Delta}H_{size}= -{\frac{1}{4}}k_a({\Delta}a)^2+{\frac{1}{2}}k_c({\Delta}c)^2
\label{eq8}
\end{equation}

The term ${\Delta}H_{charge}$ is a function of charge distribution ${\rho}(r)$, and the exact form could be exceedingly complex. However, it can be simplified in the 1144 scenario by noticing that there are limited choices of ${\rho}(r)$. Neglecting the minor distinctions due to the lattice constants, the charge distributions of, for instance, CaKRu$_4$P$_4$ and SrRbRu$_4$P$_4$ should be similar, both of which can be represented by the sketch of Fig.~\ref{f1}(c). Such resemblance helps classify 1144-phosphides into 13 groups, by the criterion that the in-group compounds should be composed of the same TM and cation valence (Sec. 5 of SI \cite{SI}). For example, BaRbFe$_4$P$_4$ belongs to the same group as CaKFe$_4$P$_4$; however, BaRbRu$_4$P$_4$ (for distinct TM) and LaKFe$_4$P$_4$ (distinct cation valence) do not. In each group, the ${\rho}(r)$ are considered identical. The limited options for ${\rho}(r)$ reduce the functional ${\Delta}H_{charge}[{\rho}(r)]$ to a function that takes 13 distinct values (Table~\ref{tab0}). Within each group, we assume the ${\Delta}H_{charge}$ being a constant ${\Delta}H_{0}$ (positive means favorable for 1144). In all, there are three parameters pertaining to each group: $k_a$, $k_c$ characterizing size effects in $a$-$b$ plane and $c$-axis, respectively; and ${\Delta}H_{0}$ characterizing the charge balance effects. These parameters can be evaluated by fitting ${\Delta}H$ obtained from first-principle calculations with Eq.~\ref{eq7} and Eq.~\ref{eq8}. The results are tabulated in Table~\ref{tab0}.
\begin{table}
\caption{\label{tab:table1} Force constants $k_a$ $k_c$ and ${\Delta}H_{charge}$ for 13 different subgroups, obtained by fitting Eq.~\ref{eq7} and \ref{eq8} to DFT results. Units for $k$ and ${\Delta}H_0$ are meV/(atom${\cdot}$\AA{}$^2$) and  meV/atom.\label{tab0}}
\begin{ruledtabular}
\begin{tabular}{c c c c c c c c}
Ru & $k_a$  & $k_c$ & ${\Delta}H_0$ & Co & $k_a$ & $k_c$ & ${\Delta}H_0$ \\
\hline
I+II & 1675 & 1.48 & 7.17 & I+II & 2147 & -0.69 & -6.31 \\
I+III & 3395 & -0.097 & 4.34 & I+III & -1016 & -1.68 & -20.93 \\
II+II & 3816 & -1.78 & 0.39 & -- &  &  & \\
II+III & -10206 & -3.92 & -5.74 & II+III & -2632 & -4.74 & -1.81 \\
\hline
Fe & $k_a$ & $k_c$ & ${\Delta}H_0$ & Ni & $k_a$ & $k_c$ & ${\Delta}H_0$ \\
\hline
I+II & 4879 & 0.45 & 5.51 & -- & & & \\
I+III & 4370 & -0.24 & 6.34 & -- & & & \\
II+II & 1869 & -1.18 & 0.71 & II+II & 15177 & -1.43 & 0.76 \\
II+III & 393 & -2.44 & -3.86 & II+III &679  & 0.06 & 1.57 
\end{tabular}
\end{ruledtabular}
\end{table}

All the groups are yielding $|k_c|{\ll}|k_a|$ due to the common feature of 1144-phases: stronger in-plane bonding and weaker inter-layer bonding. That is understandable because covalent bonding (intra-layer) is usually stronger than ionic interaction (inter-layer). Note that negative ${\Delta}a$ and positive ${\Delta}c$ adopted in Fig.~\ref{f5} are merely a sign convention; while the signs of $k_a$, $k_c$ are not subject to choices, but are physical, determined from DFT fitting (insets of Fig.~\ref{f5}(c)(d)). The $k_a$, $k_c$ have physical meanings of force constants defined in \cite{Song} and thus should be positive. If either $k_a$ or $k_c$ is negative, it indicates unstable structures, which are seen in II+II, II+III for Fe and Ru. On the other hand, when $k_a$, $k_c$ are both positive (e.g., I+II of Fe, II+III of Ni), it suggests stable structures. Knowing $k_a$ and $k_c$, one can straightforwardly evaluate the contributions by the $a$-$b$ plane ${\Delta}H_a=-\frac{1}{4}k_a ({\Delta}a)^2$, and the $z$-axis ${\Delta}H_c=\frac{1}{2}k_c ({\Delta}c)^2$. Let us take Ru-phosphide based 1144 systems as an example. (Results for Fe-, Co-, and Ni-phosphides are found in SI \cite{SI}.) Despite $|k_c|{\ll}|k_a|$, ${\Delta}H_c$ is substantially larger than ${\Delta}H_a$ (Fig.~\ref{f5}(e) and (f)) due to the fact that ${\Delta}a{\ll}{\Delta}c$. Thus, ${\Delta}c$ serves as the primary descriptor for Ru-phosphides; this rationalizes why solely ${\Delta}c$ well accounts for the change of ${\Delta}H$ of Ru-phosphides as showed by insets of Fig.~\ref{f5}(c) and (d). (Mind that the contribution of ${\Delta}a$ is not always negligible, Sec. 4 of \cite{SI}). 

As for the charge balance, the ${\Delta}H_0$ of \textit{AB}Ru$_4$P$_4$ with I+II cations amounts to 7.17 meV/atom, comparable to the total ${\Delta}H{\sim}$10 meV/atom of 1144-phases that have been experimentally realized \cite{Song}. For II + III, ${\Delta}H_0$  is negative, battling against the positive ${\Delta}H_c$. However $|{\Delta}H_0|$  is overwhelmingly larger than ${\Delta}H_c$. Thus, the instability for II+III is due to unfavorable charge balance, rather than size effects. Note that the iso-valence Fe and Ru, which are fitted independently, yield proximate ${\Delta}H_0$ in counterparts I+II, II+II, etc. This justifies our approximation of neglecting lattice-parameter effects on ${\rho}(r)$. In contrast, the iso-valence replacement of Fe by Ru will drastically alter $k_a$ and $ k_c$. The distinct responses for size effects ($k_a$, $ k_c$) and charge effects (${\Delta}H_0$) to TM-replacement is one latent feature of the 1144 phase, for which it is beneficial to separate ${\Delta}H$ in the two terms of Eq.~\ref{eq7}. In addition, for  the case II + II, 1144-phases share the same charge distribution as 122-phases, for which ${\Delta}H$ is contributed solely by size effects and it yields ${\Delta}H_0{\sim}0$ as expected. 

In short, the entangled contributions in ${\Delta}H$ are separated into size and charge effects, which are quantitatively evaluated. The significance hierarchy for Ru-phosphide is ${\Delta}H_0{\simeq}{\Delta}H_c{\gg}{\Delta}H_a$ (Fig.~\ref{f5}(e) and (f)). For broader HC, however, ${\Delta}H_0$, ${\Delta}H_c$, and ${\Delta}H_a$ are comparable, neither of which could be safely neglected. On that account, ${\Delta}a$ ${\Delta}c$ are incomplete, although they are useful descriptors to provide screening rules for stable 1144-phases. Complexity caused by charge effects still demands quantitative evaluation.
\begin{figure*}
\includegraphics[scale=0.40]{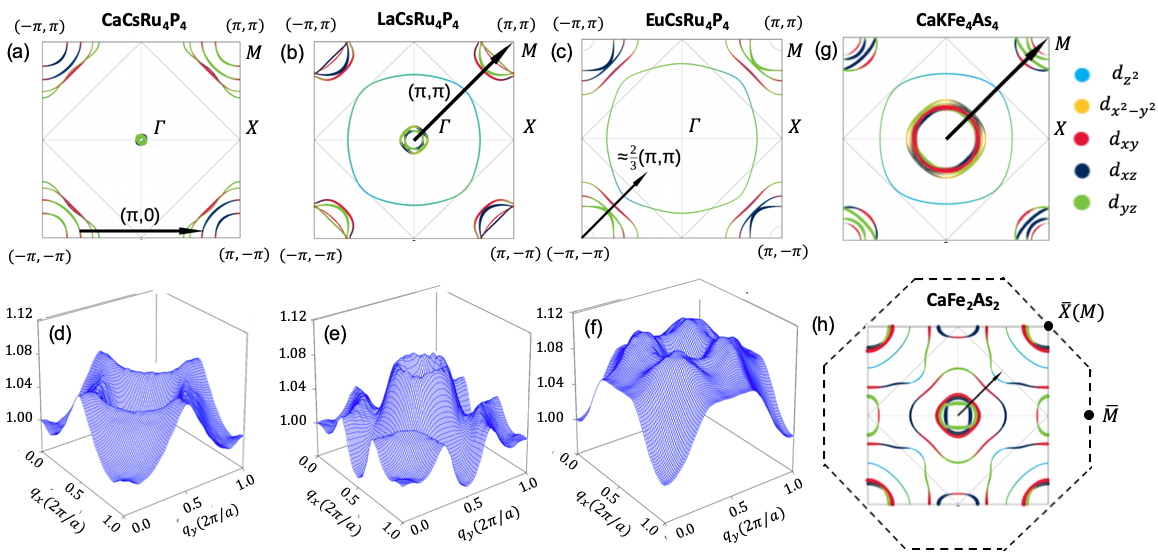}
\caption{\label{fig:epsart} (Color online) (a)-(c) FS in $k_z$=0 plane (unit for \textbf{k}: $1/a$, orbitals are colored). The main nesting vector is denoted (the nesting vector is from origin to the highest peak of ${\chi}(\textbf{q})$). (d)-(f) Non-interacting electronic susceptibility ${\chi}(\textbf{q})$ (arbitary units with $\chi(0)$ normalized to unity). (g)(f) FS of CaKFe$_4$As$_4$ and CaFe$_2$As$_2$. The B.Z. is based on the tetragonal cell (Fig.~\ref{f1}(a)) which contains 10 atoms. Dashed lines represent an alternative B.Z. based on primitive cell of 122-phases (containing 5 atoms) projected to $k_z=0$ plane (octagon), where $\bar{X}$ is equivalent to $M$ point for 1144-phases. \label{f6}}
\end{figure*}

\subsection{3C. Electronic structures}
A fascinating observation about Fe-based SC is a phase diagram that involves various degrees of freedom and symmetries. \cite{Xiang, David, DaiP} Understanding the interplay between spins, orbitals and lattices is at the heart of unraveling the origin of SC \cite{feld, Chu, Valenti} and other intertwined orders \cite{Kivelson, neumatic}. For instance, it is suggested that SC is mediated by magnetic fluctuations, which are intensified with a scenario of two-pocket Fermi surface (FS) \cite{Anderson, Schmalian}. Thus, it is interesting to examine the FS and nesting vectors of 1144-phase Ru-phosphides and compare them with 122-phase or 1144-phase Fe-arsenide.

Generally speaking, the two-pocket features include: hole-pockets near ${\Gamma}$ and electron-pockets near corner $M$ in B.Z. (Fig.~\ref{f6}). The FS pocket is cylinder-like along $k_z$ owing to the weakening band dispersion along the $z$-axis (Sec. 6 of SI \cite{SI}). That is, introducing hetero-cations impacts the FS similarly as reducing the dimension, although the periodicity along the $z$-axis remains intact. Thus, the FS could be characterized by $k_z$=0 plane (Fig.~\ref{f6}), and two-dimensional simplification for FS \cite{2Dmodel} seems plausible for the 1144-phase. On the other hand, the FS of LaRu$_2$P$_2$ strays away from the cylinder shape. This system exhibits isotropic superconductivity with $T_c$~$\sim$~4.1 K, which could be understood with phonon mediation, evidenced by a consistent e-phonon coupling strength and large density of carriers at Fermi level. \cite{Razzoli, La}. %The lack of magnetic moments for Ru \cite{Razzoli} implies a different mechanism from its Fe-As counterparts. \cite{Sasmal, Ralloy}.

%The hole-pockets at $\Gamma$ are smaller for CaCsRu$_4$P$_4$ (Fig.~\ref{f6}(a)) compared with its iso-valence peer CaKFe$_4$As$_4$ (Fig.~\ref{f6}(g)), and are even vanishing for EuCsRu$_4$P$_4$ (Fig.~\ref{f6}(c)). Seen from band structures of EuCsRu$_4$P$_4$, the bands originally at the Fermi level (FL) are suppressed ${\sim}$ 10 meV below (Sec. 6 of SI \cite{SI}), probably subjective to tunability. A common feature for Fe-arsenide is that the hole-pocket is contributed by both $d_{xz}$/$d_{yz}$, and $d_{xy}$-orbitals, as seen in Fig.~\ref{f6}(g)(h), as well as in 1111-phases, 122-phases \cite{Anderson}, and La$A$Fe$_4$As$_4$ \cite{bqs}. In contrast, in the Ru-P 1144-phase, the $d_{xy}$ disappears from the pockets at ${\Gamma}$ (Fig.~\ref{f6}(a)-(c)). Notice that, in between the two pockets, there is another FS sheet contributed by $d_{yz}$, which is also seen in La$A$Fe$_4$As$_4$ \cite{bqs}, but absent for CaCsRu$_4$P$_4$. The distinction is mainly due to the so-called half-collapsed tetragonal phase (h-CT)\cite{Xiang}, i.e., the pnictogen atoms in neighboring layers start to form $p_z$ bonds \cite{cCa} and push the bonding and anti-bonding states away from the FS.

In 1144-phases, $d$-electrons of Ru$^{2+}$ occupy two spins equally, exhibiting zero magnetic moments, which suggests a ground state of diamagnetism or paramagnetism. In contrast, in Fe-asenide, Fe has a fraction of effective magnetic moments \cite{MeierNat, Anderson, Sasmal}. For instance, KCaFe$_4$As$_4$ bears$\sim$0.2 ${\mu}_B$, and forms long range spin vortex ordering after Ni- or Co-doping \cite{MeierNat}. Note that even paramagnetic phases might exhibit AFM magnetic fluctuations \cite{Yuji}, which is a relic of long-range ordering. Thus, it is unclear whether the absence of long-range order might be changed with doping or other tunings. To shed light on this, we estimate the non-interacting electronic susceptibility \cite{Hubbard, Kiq} (Fig.~\ref{f6}(d)-(f)). 
\begin{equation}
\begin{split}
{\chi}(\textbf{q})=&-\frac{1}{V}{\sum}_{\textbf{k}, n, n'}\frac{f_n(\textbf{k}+\textbf{q})-f_{n'}(\textbf{k})}{{\varepsilon}_n-{\varepsilon}_{n'}+i{\delta}}\\
&{\times}{\langle}\textbf{k},n|e^{-i\textbf{q}{\cdot}\textbf{r}}|\textbf{k}+\textbf{q},n'{\rangle}{\langle}\textbf{k}+\textbf{q},n'|e^{i\textbf{q}{\cdot}\textbf{r}}|\textbf{k},n{\rangle}.
\end{split}
\label{eq9}
\end{equation}
The $f_n$ and ${\varepsilon}_n$ are Fermi distribution and $n$th band energies. The $V$ is volume and $|\textbf{k},n\rangle$ stands for Bloch states. In our calculation, the transition matrix ${\langle}\textbf{k}+\textbf{q},n'|e^{i\textbf{q}{\cdot}\textbf{r}}|\textbf{k},n{\rangle}$ is taken to be 1. Thus, $\chi(\textbf{q})$ is the non-interacting susceptibility under the constant phase approximation, and can shed light on the actual susceptibility.  

The main nesting vector (black in Fig.~\ref{f6}(a)-(c)) varies with cation combinations. CaCsRu$_4$P$_4$ displays a commensurate (0, ${\pi}$) or (${\pi}$, 0), reflecting the nesting between pockets at two neighboring $M$. LaCsRu$_4$P$_4$ shows a FS similar to CaKFe$_4$As$_4$ (Fig.~\ref{f6}(g)), and exhibits the same vector (${\pi}$, ${\pi}$), due to the nesting between ${\Gamma}$ and $M$. It has been found that below certain value of Pn-height the nesting vector (0, ${\pi}$) will change to (${\pi}$, ${\pi}$) \cite{Nest}. In our case, the Pn-height (the minimum) for CsCa is 1.13 \AA{} and CsLa is 1.11 \AA{}, which endorses the tendency found in the scenario of 11-phase \cite{Nest}. In EuCsRu$_4$P$_4$, the pocket at ${\Gamma}$ is absent, but has an additional FS sheet, for which ${\chi}(q)$ reach cusps at incommensurate positions $\frac{2}{3}({\pi}, {\pi})$ (Fig.~\ref{f6}(f)). In comparison, 122-phase Fe-arsenide, such as CaFe$_2$As$_2$ (Fig.~\ref{f6}(h)), exhibits a vector $\frac{1}{2}({\pi}, {\pi})$. Note that two choices of first B.Z. have been commonly used for 122-phases \cite{David}: body-centered tetragonal (two formula units) and primitive. The latter one is more convenient to compare with the 1144-phase. The two choices are compared in Fig.~\ref{f6}(h).

\subsection{3D. Crystal structures}
\begin{figure}
\includegraphics[scale=0.55]{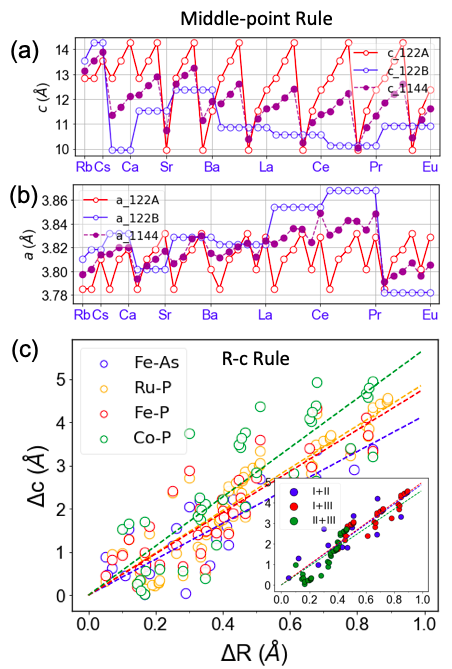}
\caption{\label{fig:epsart} (Color online) (a) The $c$-lattice parameters (calculation) for various 1144-phases (solid) and its parent phases (hollow). The solid spot is on the average of the two hollow spots. The full list of $AB$ (from left to right): KRb, KCs, RbCs, KCa, RbCa, CsCa, KSr, RbSr, CsSr, CaSr, KBa, RbBa, CsBa, CaBa, SrBa, (K, Rb, Cs, Ca, Sr, Ba)Ln (Ln=La, Ce, Pr, Eu). For brevity, only $B$ is listed on section boundaries. (b) The middle-point rule fails for $a$-lattice parameters, as considerable deviations are observed for CsCa, RbBa, CsBa, etc. (c) Linear fitting of ${\Delta}c$ with ${\Delta}R$. The slopes of different 1144-systems are quite close: \textit{AB}Fe$_4$As$_4$: 4.16, \textit{AB}Ru$_4$P$_4$: 4.90, \textit{AB}Fe$_4$P$_4$: 4.78, \textit{AB}Co$_4$P$_4$: 5.69. The inset shows the fitting with grouping cation combinations, not distinguishing TM. The slopes are 4.91, 5.00, and 4.61 for I+II, I+III and II+III cations, respectively.\label{f7}}
\end{figure}

The crystal parameters (e.g., pnictogen angles and lattice constants) are interesting to study \cite{Angle, StrPro, PnHa, PnHb, Broholm}. We discuss them from three aspects. The first aspect regards two general rules about 1144-phases, namely \textit{middle-point rule} and $R$-$c$ \textit{rule}. The middle-point rule relates the lattice constants of the 1144-phase with those of its parent phases. That is, the $c$-lattice parameter of 1144-phase $c_{1144} = \frac{1}{2}(c_{122}^A + c_{122}^B)$, where $c_{122}^A$ and $c_{122}^B$ are the $c$-lattice parameters of the two 122-phases (Fig.~\ref{f7}(a)). For example, KCaRu$_4$P$_4$ is at the middle point of KRu$_2$P$_2$ and CaRu$_2$P$_2$. Remarkably, the rule generally applies to 1144-phases of phosphides and arsenides, and of various TM. However, the rule is only for the $c$-lattice, and considerable deviations are observed for $a$-lattice parameters (Fig.~\ref{f7}(b)). This finding suggests that complexity in 1144-systems (probably also for general hetero-crystals) could be segregated, i.e., the $z$-axis is the easy part and complexity mainly arises from the $x$-$y$ plane. The other rule, $R$-$c$ rule, links crystal features with atomic information: the mismatch of the $c$-lattice parameters of two 122-phases ${\Delta}c=|c_{122}^A-c_{122}^B |$ is proportional to the mismatch of atomic radius ${\Delta}R=|R_A-R_B|$, where $R_A$ and $R_B$ are atomic radii of cation elements (metallic radius \cite{radius}). Noteworthy, a linear correlation is satisfactorily yielded by the fitting of various TMs (Fig.~\ref{f7}(c)) and cation combinations (inset of Fig.~\ref{f7}(c)); again, such a correlation does not hold for $a$-lattice parameters. 
The linearity could be partically rationalized by the intuitve picture of imagining the cation as a ``hard ball". Since each unit cell contains two layers of cations, ${\Delta}c$ should be linear with two times the diameter difference, i.e., 4${\cdot}{\Delta}R$. However, the average slope is around 5${\cdot}{\Delta}R$, somewhat deviating from the simple intuitive picture. The $R$-$c$ rule is well verified by existing experimental data \cite{Iyo, Song} and a comparison is given in Sec. 8 of SI \cite{SI}. This reconciles the distinct descriptors used in previous studies \cite{Iyo, Song}, i.e., ${\Delta}R$ and ${\Delta}c$ are equivalent. 

%Using ionic or atomic radius will not make a qualitative difference.

The second aspect concerns the strong asymmetry in structural parameters. Since the 1144-phase is the stacking of two 122-phases, we may wonder whether the TM-plane separation (inter-layer distance) in the 1144-phase remains the same as in the 122-phase. We define ${\Delta}d=d_{TM}^{1144}-d_{TM}^{122}$ to describe its displacement direction and amplitude (Sec. 7c of SI \cite{SI}). For Ru-phosphides, the TM-planes tend to move toward the higher-valence cation and away from the lower-valence. For instance in CaKRu$_4$P$_4$, the separation between TM-layer and  Ca$^{2+}$ shrinks, while the separation between TM-layer and K$^+$ increases. On the contrary, for Co, the TM-plane will displace toward the lower-valence side. Ni- and Fe-1144-phases display both tendencies. \cite{SI} Whichever direction it will move, the displacement magnitudes $|{\Delta}d/d_{122}|{\sim}5\%$-$10\%$(Sec. 7 of \cite{SI}) are considerably larger than the structural change that is achievable by applying pressure. Besides, pnictogen (Pn) heights and Pn-TM-Pn bonding angles also undergo large changes. The large change does not happen in solid solution phase, for each cation layer takes equal amounts of charge, unlike $A$, $B$ are living separately in 1144-phases, which intensely breaks the glide symmetry and affects superconductivity. \cite{Angle, PnHa, PnHb}. 

The third aspect is regarding derivative phases, known as the half-collapsed phase (h-CT). In general, derivative phases can arise from particular architectures; for example, molecule folding emerges from long-chain architectures. h-CT refers to an abrupt collapse of the lattice constant $c$ at certain external pressure, first observed in KCaFe$_4$As$_4$ \cite{Collapse}. The collapse happens in two steps: the first-step in Ca-layers, the second in K-layers \cite{TwoTr}; so h-CT refers to such a state where one layer collapses and the other does not. h-CT is peculiar for HC as it is determined by two general features: layer-building and distinct compressive capacities of hetero-cation layers\cite{Xiang, Borisov, TwoTr}. h-CT is a consequence of $A$-$B$ symmetry breaking, and also grants a method of tuning asymmetry, which leads to, for instance, suppression of SC \cite{LoseSC}. Forming the collapsed phase is related to the energy valley of a double-minimum \cite{Pbond}. In the language of bonding, it requires the Pn-Pn bonding to be strengthened and Pn-TM bonding to be weakened. Therefore, there exists a critical distance $d_P$(defined in Fig.~\ref{f1}(a)), below which the collapsed phase will be formed. The critical $d_P$ is found to be 2.8$\sim$3.0 \AA{} for Fe-arsenide 1144-phases, which is achievable with a moderate external pressure \cite{Borisov}. Noteworthy, Ru-phosphides 1144-systems exhibit smaller $d_P$ and thereby may form h-CT with no or very little pressure, easier than Fe-arsenide. The critical $d_P$ for phosphide is about 2.5 \AA{} (Sec. 7e of \cite{SI}).

\section{4. Experimental results}
In this section we present our design and discovery of CaKRu$_4$P$_4$, a new example of a 1144-type compound and a proof of principle of our ability to identify and create new complex phases.  CaKRu$_4$P$_4$ was chosen as a target phase based on the simultaneous large difference in the $c$-lattice parameter values and small difference in $a$-lattice parameter values exhibited by the KRu$_2$P$_2$ and CaRu$_2$P$_2$ ternary compounds.  In addition, from a technical point of view, we had already mastered the simultaneous use of K and Ca as part of our efforts to grow CaKFe$_4$As$_4$. \cite{MeierPRM} We first synthesized KRu$_2$P$_2$ and CaRu$_2$P$_2$ as precursor compounds and then combined them to create the quaternary CaKRu$_4$P$_4$.  We want to emphasize that this synthesis truly was based on designing the compound, based on our understanding of the requirements needed to stabilize the 1144 structure, and then implementing this design in our synthesis.

\subsection{4A. Synthesis procedure}
Polycrystalline CaKRu$_{4}$P$_{4}$ samples were growth by solid state reaction. CaRu$_{2}$P$_{2}$ and KRu$_{2}$P$_{2}$ polycrystalline samples were synthesized first as precursors which were prepared by putting lumps of potassium metal (Alfa Aesar 99.95\% metals basis) or calcium metal pieces (Ames Laboratory, Materials Preparation Center (MPC) 99.9\%), ruthenium powder (Alfa Aesar 99.95\% metals basis) and lumps of red phosphorus (Alfa Aesar 99.999\% metals basis) into an alumina crucible that was arc-welded into a Ta tube and finally sealed into an amorphous silica ampoule. Because of the high vapor pressure of phosphorus, the temperature of furnace was ramped slowly, dwelling at 230\textcelsius\ and 500\textcelsius\ for 1 and 10 hours respectively, before being increased to 900\textcelsius\ and reacting for 96 hours. After cooling down to room temperature the initial reaction product was ground into powder and pressed into a pellet by tungsten carbide die press in argon filled glove-box. The pressed pellet was sealed in the manner described above and reacted at 900\textcelsius\ for another 96 hours.  This process was repeated until phase pure KRu$_{2}$P$_{2}$ polycrystalline samples were synthesized. The similar process was repeated several times for CaRu$_{2}$P$_{2}$ until the phase was almost pure.   Finally a pellet was pressed from a 1:1 molar ratio of CaRu$_{2}$P$_{2}$ and KRu$_{2}$P$_{2}$.  This pellet was heated to 850\textcelsius\ and kept there for 95 hours after which the pellet was cooled to room temperature, ground, pressed and resealed.  This procedure was repeated until the CaKRu$_{4}$P$_{4}$ phase (88\% purity) was formed. Higher purity is thought to be possible with proper adjustment of  precursors ratio before reaction.

\subsection{4B. Structural analysis}
Powder x-ray diffraction measurement were carried out on finely ground CaRu$_{2}$P$_{2}$ and KRu$_{2}$P$_{2}$ using a Rigaku MiniFlex II powder diffactometer with Cu K$\alpha$ radiation ($\lambda$ = 1.5406 \AA{}) with refinements done by GSAS.{\cite{1}} The PXRD data of CaKRu$_{4}$P$_{4}$ was recorded at room temperature on a PANalytical X-Pert Pro Diffraction System with Co K$\alpha$ radiation ($\lambda$ = 1.78897 \AA{}). Powdered sample was evenly dispersed on a zero-background Si-holder with the aid of a small quantity of vacuum grease. Intensities were collected for the 2$\theta$ range from 15$\degree$ to 110$\degree$ in step sizes of 0.02$\degree$ and dwell times of 300 s for each step. FullProf Suite program package was used for Rietveld refinement of the crystal structures {\cite{2}}.

\begin{table}
\caption{X-ray data refinement including space group, formula units/cell and refined lattice parameters for
CaRu$_2$P$_2$ and KRu$_2$P$_2$. \label{table0}}
\begin{ruledtabular}
\begin{tabular}{c c c c c c}
CaRu$_2$P$_2$ & \textit{I4/mmm} & $a$: 4.044(4) & $c$: 9.834(1) & & \\
\hline
Atom(Mult) & $x$ & $y$ & $z$ & Uiso & Occ. \\
\hline
Ca (2) & 0 & 0 & 0 & 0.00361 & 1.0 \\
Ru (4) & 0 & 1/2 & 1/4 & 0.00357 & 1.0 \\
P  (4) & 0 & 0 & 0.368(7) & 0.00300 & 1.0 \\
\hline
KRu$_2$P$_2$ & \textit{I4/mmm} & $a$: 4.016(7) & $c$: 12.356(9) & & \\
\hline
Atom(Mult) & $x$ & $y$ & $z$ & Uiso & Occ. \\
\hline
K (2) & 0 & 0 & 0 & 0.00869 & 1.0 \\
Ru (4) & 0 & 1/2 & 1/4 & 0.00160 & 1.0 \\
P  (4) & 0 & 0 & 0.340(5) & 0.00437 & 1.0 \\
\end{tabular}
\end{ruledtabular}	
\end{table}

\begin{table}
\caption{X-ray data collection, refinement and residual parameters for the experimental sample including space group(SG), formula units/cell (\textit{Z}) and refined lattice parameters for each of the constituent phase. \label{table1}}
\begin{ruledtabular}
\begin{tabular}{c c c c}
 & CaKRu$_4$P$_4$ & RuP & CaRu$_2$P$_2$ \\
\hline
SG, \textit{Z} & \textit{P4/mmm}(123), 1 & \textit{Pnma}(62), 4 & \textit{I4/mmm}(139),2 \\
$a$(${\AA}$) & 4.048(1) & 5.524(1) & 4.050(1) \\
$b$(${\AA}$) &  & 3.158(1) & \\
$c$(${\AA}$) & 10.913(1) & 6.128(1) & 9.756(1) \\
$V$(${\AA}^3$) & 178.85(3) & 106.95(2) & 160.34(3) \\
Radiation &  & Co $K_a$=1.78901 & \\
2${\theta}$ &  & 15-110$^{\circ}$ & \\
No. of data &  & 5679 & \\
Uniq. data & 152 & 112 & 70 \\
No. of Var. &  & 31 & \\
residuals & $R_B$=0.12 & $R_B$=0.22 & $R_B$=0.17
\end{tabular}
\end{ruledtabular}	
\end{table}

\begin{table}
	\caption{Atomic Coordinates for CaKRu$_4$P$_4$, Isotropic Displacement Parameters (\AA{}$^{2}$), and Site Occupancies for the main phase\label{table2}}
\begin{ruledtabular}
\begin{tabular}{c c c c c c c c}
atom & position  & Symm. & $x$ & $y$ & $z$ & Biso & Occ. \\
\hline
Ca & 1\textit{a} & 4/\textit{mmm} & 0 & 0 & 0 & 0.83(1) & 1.0  \\
K & 1\textit{d} & 4/\textit{mmm} & 1/2 & 1/2 & 1/2 & 1.25(1) & 1.0 \\
Ru & 4\textit{i} & 2\textit{mm} & 0 & 1/2 & 0.217(1) & 1.22(1) & 1.0 \\
P1 & 2\textit{g} & 4\textit{mm} & 0 & 0 & 0.340(1) & 1.92(1) & 1.0 \\
P2 & 2\textit{h} & 4\textit{mm} & 1/2 & 1/2 & 0.107(1) & 0.28(1) & 1.0
\end{tabular}
\end{ruledtabular}	
\end{table}

\begin{figure}
\includegraphics[width=8.0cm]{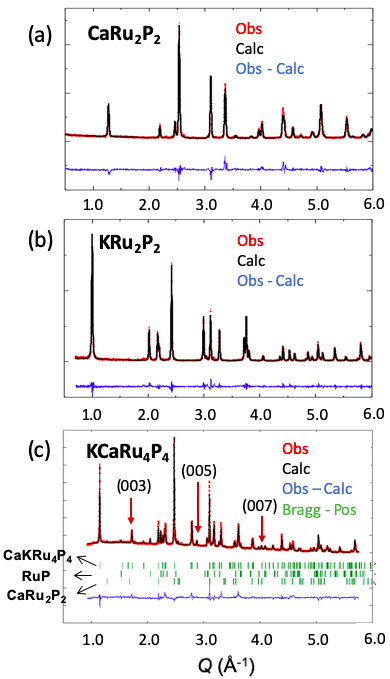}%	
\caption{(a)(b)Powder XRD pattern and Refinement of CaRu$_2$P$_2$ and KRu$_2$P$_2$. (c) PXRD and Rietveld refinement results for the CaKRu$_4$P$_4$. (003), (005) and (007) peaks are indicated by the red arrows. The observed profile is indicated by red circles and the calculated profile by the solid line. Bragg peak positions are indicated by vertical tics, and the difference diffractogram is in blue. \label{figure8}}
\end{figure}

The powder XRD pattern and Rietveld refinement for KRu$_{2}$P$_{2}$, CaRu$_{2}$P$_{2}$ and CaKRu$_{4}$P$_{4}$ are shown in Fig.~\ref{figure8}.  X-ray refinement for CaRu$_2$P$_2$ and KRu$_2$P$_2$ are shown in Table \ref{table0}.  The small difference in a-axis lattice parameters (4.044 versus 4.016 \AA{}) and the large difference in c-lattice parameters (9.834 versus 12.356 \AA{}) for CaRu$_2$P$_2$ and KRu$_2$P$_2$ respectively are clearly seen.  For the CaKRu$_4$P$_4$ sample, the X-ray data collection specifications, refined lattice parameters along with residual factors for all three phases are given in Table \ref{table1}, whereas the refined coordinates and isotropic displacement factors are presented for the main phase in Table \ref{table2}.  The clear appearance of peaks with h + k + l = odd in Fig.~\ref{figure8}(c). is the qualitative indication that the CaKRu$_4$P$_4$ phase has formed \cite{Iyo}. 

For the CaKRu$_{4}$P$_{4}$ refinement, first the zero-point shift and scale factor were refined. Next the lattice constants were refined, followed by the background parameters which involved a gradual increase of the polynomial used in the fit up to the sixth order. Consequently, the positional parameters are refined, beginning with the ruthenium atoms and followed by all phosphorus atoms. This is followed by refinement of the isotropic thermal parameters for all atoms at fixed 100\% occupancies. Then the occupancy parameters are unfixed and refined together with thermal parameters. This step confirms 1:1:4:4 stoichiometry of the phase, as well as the Ca/K segregation to their separate unique sites, since minimal fluctuation ($<$1 \%) of the occupancies is revealed at virtually same thermal parameters. Afterwards, profile-shape parameters are included with a peak-asymmetry correction for 2$\theta<40\degree$. The final refinements with 31 parameters, including preferential alignment, and 5,679 data points were performed. Phase analysis of the data in  Fig~\ref{figure8}(c) gives $\sim$ 88\% of CaKRu$_{4}$P$_{4}$ estimated with $\sim$10\% CaRu$_{2}$P$_{2}$ and $\sim$ 2\% RuP after the final reaction. The existence of other phases may be due to imbalance of moles of CaRu$_2$P$_2$ and KRu$_2$P$_2$ during reaction.
%air sensitive of KRu$_{2}$P$_{2}$. Some of the KRu$_{2}$P$_{2}$ may decompose to binary before react with CaRu$_{2}$P$_{2}$. 

\section{5. Discussion}
\begin{figure}
\includegraphics[scale=0.42]{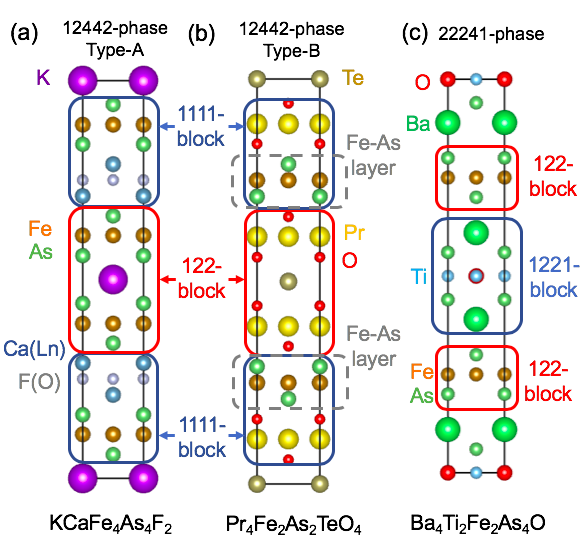}
\caption{\label{fig:epsart} (Color online) Examples of constructing hetero-crystals. (a)(b) Two types of 12442-phase. Differences: type A: 4 layers of Fe-As and 2-layers of composite cation layers (Ca-F)$^{+}$ per unit cell; type B: 2 layers of Fe-As and 4 layers of cation layers (Pr-O)$^{+}$ (c) 22241-phase.\label{f9}}
\end{figure}

Here we discuss hetero-crystals beyond 1144-phases. The so-called 12442-phase (e.g., K(Cs)Ca$_2$Fe$_4$As$_4$F$_2$, \cite{C12442, CaoNew} Pr$_4$Fe$_2$As$_2$TeO$_4$ \cite{C42214}) is obtained by stacking two 1111-phases on either side of one 122-phase (Fig.~\ref{f9}(a)). Since Ln-O or Ca-F layers effectively serve as 1+ cations, it can be considered as alternative stacking of elemental cations Ca$^{2+}$ and composite cations (Ln-O)$^{+}$ or (Ca-F)$^{+}$; the TM-layer is (Fe-As)$^{1.5-}$. Notice that the 12442-phase has a second stacking type (Fig.~\ref{f9}(b)), with Fe-As and La-O interchanging their roles. We may still think of it as 1111+122, but an alternative viewpoint is the Fe-As layer being sandwiched by two dislocated layers of 122-phases. The latter viewpoint is more insightful because the Fe-As layer is the playground for SC and magnetism, while Ca or La-O altogether merely serve as electron donators. A particular stacking of the two could be achieved by selecting proper precursors \cite{C12442, C42214}. The 12442-phase implies that the cation could be a composite one, like (Ln-O)$^{+}$, expanding the pool for cations, as well as allowing extra tuning parameters: the stacking type (distinction of Fig.~\ref{f9}(a) and (b)). Other examples include the 22241-phase (Fig.~\ref{f9}(c)) \cite{C22241}, which contains alternative stacking of 1221-phases and 122-phases. Interestingly, it builds in oxide motifs, which are another big familiy of parent compounds. Recently, 112-phase nickelate SC attracted much attention \cite{NiX, NiL}, which belong to the same structure family of cuprate SC \cite{Cufmly} and is a limiting case $n\rightarrow{\infty}$ for the series of compounds La$_{n+1}$Ni$_n$O$_{2n+1}$ \cite{NiOfmly}. In 112-phases, cations RE or IIA atoms are sandwiched by NiO$_2$ layers. We have seen many hetero-crystals in Fe-SC and it is interesting to examine Ni- or Cu-based SC.

Here, we shall revisit hetero-crystal's definition and compare it with some similar-looking compounds. (1) Layered structure ((ii) in Sec. 2) is an indispensible defining feature, i.e., bonding along z-axis should be weaker. On top of that, quasi-2D FS can form, nesting and unconventional SC are established. On the other hand, structures, such as pyrochlores or fluorite, \cite{PFtr} are 3D corner-sharing networks, featuring very different properties. Since they lack “layers”, the notions developed here, such as charge balance between layers, layer-stacking distortions, and corresponding descriptors may not apply. Thus, even with similar looks ($A$-$B$ ordering), they are different types. (2) Asymmetry is another crucial characteristic ((iii) in Sec. 2). It does not exclude such situation as, for instance, $AABAAB$-stacking, because the double-layer $AA$ could be viewed as a single combined layer and asymmetry still occurs at the interface between $AA$ and $B$. On the same account, it is acceptable for having a certain amount of stacking faults or disorder (e.g., the $A$-layer contains 10\% $B$), as long as imperfections will not eliminate all the asymmetry. Conceptually, hetero-crystal is $not$ equated with disorder free; in fact, disorder could be introduced on purpose as seen in Ni- or Co-doped CaKFe$_4$As$_4$ \cite{NoNeum}. In that case, disorder enter through TM-layers not cation layers, and cation polarization maintains. The asymmetry can be characterized by defining $\varepsilon=(d_{TM}(A)-d_{TM}(B))/{\Delta}d_0$ or ${\varepsilon}=({\alpha}(A)-{\alpha}(B))/{\Delta}{\alpha}_0$, where ${\Delta}d_0$ and ${\Delta}{\alpha}_0$ are normalization factors. Then, the definition refers to a finite regime of $\varepsilon$, rather than on the single perfect point ${\varepsilon}=1$. (3) HC hybridizes two \textit{stable} phases/motifs ((iv) in Sec. 2). However, compounds like NaYbO$_2$ \cite{NaYbO2} cannot be regarded as NaO+YbO since each individual is unstable. Thus, NaYbO$_2$ is one phase with $A$-$B$ ordered cations, rather than $A$-$B$ ordered hybrids of two phases. Note that in the context of, for instance, pyrochlores-fluorite transition, the initial phase is ``parent” and the resultant phase is ``derivative”. Here, ``parent phases" have different meanings, because firstly they must be in pair and then be hybridized. Without the \textit{two} parent phases, pseudo-binary system might not be guaranteed, entropy formula Eq.~\ref{eq5} will become groundless. 
\begin{figure}
\includegraphics[scale=0.42]{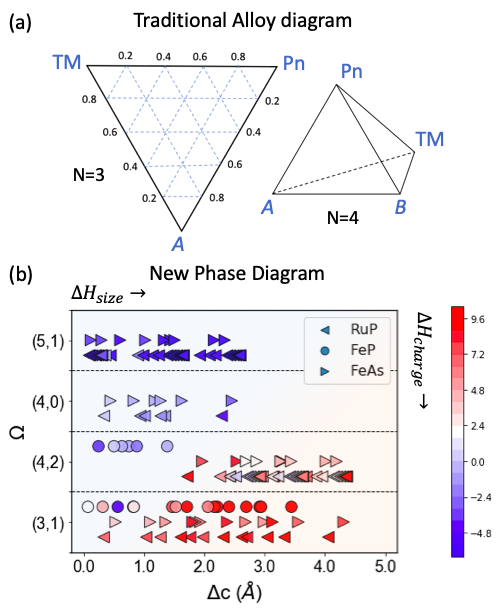}
\caption{\label{fig:epsart} (Color online) (a) Traditional phase diagram with $N=3$ (triangle) and $N=4$ (tetrahedron). The complexity of phase diagram will quickly increase with ingredient number $N$. Stable phases only sparsely distribute, leaving phase separation in most regions. \cite{Gaskell} (b) Formation enthalpy of 1144-phases $AB$Ru$_4$P$_4$, $AB$Fe$_4$P$_4$, $AB$Fe$_4$As$_4$ (color scales, meV/atom) vs collective descriptors as coordinates, with which the phase diagram could include compounds with flexible compositions and concentrations. It has a clear separation betwen the stable (red) and the unstable (blue) and is a generalization of phase diagram in \cite{Iyo}.\label{f10}}
\end{figure}

%One growth method is high-temperature solution \cite{MeierPRM, MeierNat}. Appropriate compositions and protection to the active metal vapor are important factors to succeed. \cite{MeierPRM}. In addition, growing single-crystals requires a path to the solidification window in the temperature-composition space. A second common approach is solid state reaction \cite{Iyo, CaoRbEu}, for which the precursor choice is important. Elements e.g., K, are active and the heats released by them prevent getting the desired reaction. Our strategy is forming parent phases first and react their powders. It is also possible to use hydrides BaH$_2$ and KH as source of the active metals \cite{KH}.

Hetero-crystals allude to a fundamental view of exploring complex alloys, as reflected by a new type of phase diagram (Fig.~\ref{f10}). A traditional phase diagram is coordinated by information of an individual atom (e.g., elemental species and concentrations) and becomes cumbersome when $N>3$ (Fig.~\ref{f10}(a)). We introduce a different set of coordinates: ${\Delta}c$ (or ${\Delta}a$) defined by Eq.~\ref{descriptor_a}, \ref{descriptor_c}, and a tuple ${\Omega}=(n_A+n_B, |n_A-n_B|)$, where $n_A$, $n_B$ are valence electron numbers for cations $A$ $B$.  For example, KCaRu$_4$P$_4$ has $n_A=1$(K), $n_B=2$(Ca), thus ${\Omega}=(3,1)$. These coordinates rely on the whole collection of involved atoms, rather than each individual, thus referred to as ``collective". Kind of the idea of using transformed coordinates $q$ to describe collective modes; meanwhile, the information remains unchanged, since ${\Delta}c$, ${\Delta}a$, ${\Omega}$ are ultimately from the individual information.
\begin{table}
\caption{\label{tab:table2} Formation enthalpy for stable 1144-systems. Units for ${\Delta}a$ (${\Delta}c$) and ${\Delta}H_0$ are ${\AA}$ and  meV/atom, respectively\label{table4}}
\begin{ruledtabular}
\begin{tabular}{c c c c c c c c}
Ru-P & $|{\Delta}a|$  & $|{\Delta}_c|$ & ${\Delta}H$ & Fe-P & $|{\Delta}a|$ & $|{\Delta}c|$ & ${\Delta}H$ \\
\hline
KCa & 0.026 & 2.623 & 14.98 & KEu & 0.009 & 1.806 & 8.14 \\
RbCa & 0.022 & 3.328 & 15.44 & RbEu & 0.007 & 2.380 & 8.68  \\
CsCa & 0.033 & 4.057 & 18.04 & CsEu & 0.009 & 3.130 & 8.87 \\
KLa & 0.009 & 1.721 & 10.04 & KLa & 0.030 & 1.966 & 7.58 \\
RbLa & 0.005 & 2.426 & 10.19 & RbLa & 0.028 & 2.540 & 8.50 \\
CsLa & 0.016 & 3.155 & 11.79 & CsLa & 0.012 & 3.290 & 9.40 \\
KEu & 0.008 & 1.611 & 12.90 & & & & \\
RbEu & 0.004 & 2.316 & 14.19 & & & & \\
CsEu & 0.015 & 3.045 & 15.17 & & & &
\end{tabular}
\end{ruledtabular}
\end{table}

As such, one obtains a new phase diagram (Fig.~\ref{f10}(b)) of reduced dimensions, which can embody entries with flexible chemical ingredients, such as Fig.~\ref{f10}(b) that has included Fe-arsenide, Ru- and Fe-phosphide. Descriptors ${\Delta}c$, ${\Omega}$ do not depend on the number of constituents, thus complexity will not rely on the compounds being ternary, quaternary or beyond. In addition, the separating between stable and unstable structures is sharp, which reflects the fact that ${\Delta}c$ and ${\Omega}$ have captured two universal factors for hetero-crystals: size and charge effects. That is, $x$-axis ${\Delta}c$ is in charge of ${\Delta}H_{size}$ and $y$-axis ${\Omega}$ in charge of ${\Delta}H_{charge}$.

Traditional phase diagram is an efficient apparatus for SS that forms when entropy dictates. The hetero-crystal is an abnormality emerging out of the SS region, thus invokes new principles and methodology. The new paradigm changes the viewpoint about composite materials, which were based on combining multiple elementary atoms and tagged by the number of species, such as binary, ternary. In contrast, a hetero-crystal is always viewed as a ``binary" system that combines two building units; but these units could be composite objects. In other words, for complex alloys, it is beneficial (also necessary) to introduce intermediate structure units in between the elementary level and the final crystal. Just like we consider biological structures as aggregation of protein molecules rather than elementary atoms. Parent phases or motifs play the role of the building units. Accordingly the attributes must be modified: ${\Delta}c$, ${\Delta}a$ are derived from parent phases, known as collector descriptors. The strategy for synthesizing will also be modified. A common procedure to discover new compounds is choosing a structure prototype and substituting certain atom with another of similar individual properties, like radius or electronegativity. It shifts the focus to studying the matching (or mismatch) of parent phases and motifs, which might lead to numerous hidden matter states.

At last we discuss the experimental realization. The compound La$A$Fe$_4$As$_4$ ($A$=K, Rb, Cs) is predicted to be stable\cite{bqs}, but its synthesis is still lacking. This is probably due to the parent phase LaFe$_2$As$_2$ being meta-stable, a significant difference from phosphides examined in present work. However, LaFe$_2$As$_2$ is recently synthesized \cite{LaFeAs}. Using LaFe$_2$As$_2$ as precursors renews the hope. This ``failure" example also suggests the importance of stable parent phases (condition (iv) in Sec. 2). 

\section{6. Summary}
In summary, we introduce a class of $A$-$B$ hetero-layer intermetallic crystals and address the following specific questions. What are they? Broadly existing or not? What are the main features? What influences or insights will they bring?

It is inspired by simple intuition: make the random substitution in metallic alloys into an ordered $A$-$B$ stacking; in other words, replace the layers in semiconductor hetero-structures by some metallic ingredients, such as TM-layers and cations. To make the idea precise, four definitions are given (Sec. 2): (i) bulk materials, (ii) cation layers plus TM-layers, (iii) $A$-$B$ stacking and reduced symmetry, (iv) formed with two parent phases. 

Hetero-crystals may exist broadly, as seen in Fig.~\ref{f5}, \ref{f10}. By computational searching in phosphides of 1144-phase (TM=Fe, Ru, Co, Ni), we found a series of stable structures at ambient temperatures and pressures, particularly for Fe- and Ru-phosphides. The most promising ones are put in Table~\ref{table4} for a quick survey. Our prediction is supported by synthesis of high-purity KCaRu$_4$P$_4$. The mechanism is the battling of enthalpy and configuration entropy. The enthalpy rises from size effects and charge balance, which could be characterized by two descriptors ${\Delta}c$ and ${\Omega}$ (Fig.~\ref{f10}).

For crystal features, the TM-layers commonly has strong distortion (5\%${\sim}$10\%) compared with the parent phase. Besides, we suggest two universal latent rules: middle point rule $c_{1144} = \dfrac{1}{2}(c_{122}^A + c_{122}^B)$, and $R$-$c$ rule ${\Delta}c{\simeq}5.0{\cdot}{\Delta}R$. For electronic features, they commonly exhibit Fermi pockets, and weakening dispersion along the $z$-axis (Sec. 3C). Because of having $d$- and $f$-orbitals, they are often compounds interesting for SC, magnetism, heavy Fermions, etc.

These compounds upgrade our viewpoint about alloys, as they are always viewed as ``binary" systems that combine two parent phases; that is, the parent phase plays a role of a structure motif that bridges the atom and the crystal. Thus, It brings renewed interest of many known phases that can serve as the parent phase. Such insights also bring a new phase diagram that is coordinated by the so-called collective descriptors originating from parent phases. (Fig.~\ref{f10}). 

\textbf{Acknowledgement}. We wish to acknowledge the very helpful discussion with William Meier, Sergey Bud’ko, and Duane Johnson. Work at Ames Laboratory was supported by the U.S. Department of Energy (DOE), Office of Science, Basic Energy Sciences, Materials Science and Engineering Division including a grant of computer time at the National Energy Research Supercomputing Center (NERSC) in Berkeley. Ames Laboratory is operated for the U.S. DOE by Iowa State University under contract number DE-AC02-07CH11358. RV acknowledges support by the Deutsche Forschungsgemeinschaft (DFG, German Research Foundation) for funding through TRR 288 — 422213477 (project A05 and B05).

\end{document}